# MORSE: Multiple Orthogonal Reference Sensitivity Encoding

Oliver Josephs, Barbara Dymerska, Nadine N. Graedel, Yael Balbastre, Nadège Corbin, Martina F. Callaghan

***Abstract*—** **Parallel imaging is ubiquitous in MRI, enabling diverse applications such as ultra-high-resolution functional and quantitative imaging with greater temporal resolution or reduced scan times respectively. Successful unfolding is contingent on robust and accurate estimation of the relative coil sensitivities, which often involves computation times that preclude real-time deployment. Here we present a computationally-efficient method of robustly estimating coil sensitivities, and reconstructing under-sampled images using a data-driven regularised SENSE formalism. Our "MORSE" scheme estimates multiple sensitivities per voxel to address issues such as rapidly varying sensitivities, chemical shift artefact, or insufficient fields of view and provides a data-driven regularisation term for noise control. Exemplar structural and functional image reconstructions at 3T and 7T are presented and compared with a vendor-provided GRAPPA reconstruction as well as state-of-the-art ESPIRiT and LORAKS algorithms. MORSE consistently produced high-quality, artefact-free images with reconstruction times feasible for real-time deployment. It is flexible and robust, and made available to the community in open-source as a library of functions within the vendor-agnostic Gadgetron image reconstruction framework.**

***Index Terms*—** **image reconstruction, parallel imaging, regularised SENSE, sensitivity estimation**

## I. INTRODUCTION

IN MRI spatial encoding is comparatively slow, which motivates sampling below the Nyquist frequency to accelerate the imaging process while maintaining spatial resolution. However, this leads to aliased images with an insufficient Field of View (FoV). Receiver coil arrays enable high quality images to be constructed free of aliasing artefacts by leveraging the differential sensitivity profiles across each element/channel, i.e. parallel imaging [1-3]. Broadly speaking, calibration data are used to estimate either a local weighting kernel in k-space or a set of coil sensitivities in the image domain, which are then used to compensate for the under-sampling. There is a limit to the degree of acceleration that can be achieved because the under-sampling inherently reduces the signal-to-noise ratio (SNR) of the images. Noise can also be amplified by the image reconstruction process, particularly if the coil elements have indistinct sensitivity profiles, necessitating careful coil design [1, 4].

The SENSE algorithm, which operates in the image domain, achieves an exact and SNR-optimal reconstruction *provided* the estimates of the relative coil sensitivities are accurate [1, 5]. The simplest means of computing coil-wise sensitivity maps is with respect to a reference [5] with matched image contrast, high SNR, and coverage that extends over the entire target region with uniform or slowly varying sensitivity modulation, e.g. using a body coil image. The need for referencing can be circumvented by exploiting local correlation or a reduced rank approximation of the array coil sensitivities [6, 7]. While a wide variety of advanced image reconstruction algorithms have now been developed (see [5, 8-12] for review), computational requirements can pose a practical limit for real-time deployment, particularly for high spatial and/or temporal resolution applications.

Here we present a computationally efficient approach that permits real-time online reconstruction with moderate computing power, and without compromising image quality. Multiple Orthogonal Reference Sensitivity Encoding (MORSE) builds upon the approach of exploiting local correlations [6] to estimate accurate coil sensitivities for subsequent use in a regularised-SENSE general linear model formalism. MORSE estimates multiple orthogonal sensitivities per coil element at each voxel in the FoV to account for chemical shifts, partial volume effects, or insufficient FoV. In doing so, it addresses a number of issues that are particularly problematic at ultra-high field (UHF) strengths ($\geq$ 7T) producing high SNR images despite rapidly varying coil sensitivities and/or increased main field inhomogeneities, even with restricted FoVs, and without requiring a body coil reference, which typically isn't available at UHF.

We demonstrate utility, robustness and flexibility of MORSE with a wide variety of *in vivo* human datasets encompassing regular Cartesian under-sampling (with and without CAIPIRINHA sampling [13]) acquired with different readouts (EPI and GRE), contrasts (T2*-, T1-, PD- or MT-weighting), resolutions (0.6 to 3mm voxel lengths, isotropic and anisotropic), hardware (3T and 7T with multiple receiver coil designs) and by using both separate [14] or integrated [2] reference data to estimate the coil sensitivities. We compare

This work was supported by the Discovery Research Platform for Naturalistic Neuroimaging, funded by the Wellcome Trust (226793/Z/22/Z). (Joint first authors: O. Josephs and B. Dymerska. Corresponding author: M. Callaghan.)

O. Josephs, B. Dymerska, N. N. Graedel, and M. F. Callaghan are with The Functional Imaging Laboratory, Department of Imaging Neuroscience, UCL Queen Square Institute of Neurology, University College London, UK (e-mail: o.josephs@ucl.ac.uk, b.dymerska@ucl.ac.uk, n.graedel@uc;l.ac.uk, m.callaghan@ucl.ac.uk). Y. Balbastre is with Experimental Psychology, Division of Psychology and Language Sciences, University College London, UK (e-mail: y.balbastre@ucl.ac.uk). N. Corbin is with Centre de Résonance Magnétique des Systèmes Biologiques, UMR5536, CNRS/University Bordeaux, Bordeaux, France (e-mail: nadege.corbin@rmsb.u-bordeaux.fr).

MORSE, in terms of both computational efficiency and image quality, to vendor-native GRAPPA-based reconstructions [2, 15] and two state of the art approaches: ESPIRiT [3], openly available via the BART toolbox, and LORAKS, another open-source image reconstruction approach based on low-rank modelling of local k-space neighbourhoods [16-18].

## II. THEORY

### A. Notation

Upper case non-italic bold letters indicate tensors ($\mathbf{T} \in \mathbb{C}^{I \times J \times K}$), upper case italic bold letters indicate matrices ($\boldsymbol{M} \in \mathbb{C}^{J \times K}$), lower case italic bold letters indicate vectors ($\boldsymbol{v} \in \mathbb{C}^K$), and italic upper and lower case non-bold letters indicate scalars ($s \in \mathbb{C}$). Subscripts are used to index into a tensor, matrix or vector (e.g., $\boldsymbol{T}_k, \boldsymbol{t}_{j,k}, T_{i,j,k}\ \boldsymbol{m}_k, M_{j,k}, v_k$). Script letters indicate a set of positive integer indices ($\mathcal{N} \in \mathbb{Z}_+^K$). A superscript asterisk denotes the complex conjugate ($x^*$) while a superscript $H$ denotes the conjugate transpose ($\boldsymbol{X}^H$). All values are complex.

### B. Background

Fully sampled images, $\boldsymbol{C} \in \mathbb{C}^{N_{\text{coil}} \times N_{\text{voxel}}}$, acquired on $N_{\text{coil}}$ receiver elements from $N_{\text{voxel}}$ voxels in 3D space, correspond to the Hadamard product of the coil-wise sensitivities, $\boldsymbol{B} \in \mathbb{C}^{N_{\text{coil}} \times N_{\text{voxel}}}$, and the spatially-varying magnetisation, $\boldsymbol{\rho} \in \mathbb{C}^{N_{\text{voxel}}}$ (repeated across $N_{\text{coil}}$ rows) plus some additive measurement noise. The well-established method of Walsh *et al.* [6] estimates optimal weights for combining such coil-wise images via a matched filter approach. A set of coil-wise, voxel-specific weights (i.e. a scaled version of $\boldsymbol{B}$) is determined by performing singular value decomposition (SVD) on the outer product, $\boldsymbol{E}_r \in \mathbb{C}^{N_{\text{coil}} \times N_{\text{coil}}}$, of the coil-wise image values, $\boldsymbol{c}_r \in \mathbb{C}^{N_{\text{coil}} \times 1}$ (i.e. the column of $\boldsymbol{C}$ corresponding to voxel $r$):

$$\boldsymbol{E}_r = \boldsymbol{c}_r \boldsymbol{c}_r^H = \boldsymbol{U}_r \boldsymbol{S}_r \boldsymbol{V}_r^H \quad (1)$$

$\boldsymbol{E}_r$ is a rank one matrix with one non-zero singular value (first diagonal element of $\boldsymbol{S}_r$) for which the corresponding left singular vector (first column of $\boldsymbol{U}_r$) is a scaled version of the desired coil sensitivities at voxel $r$. A more noise-robust estimate can be obtained by assuming that each coil sensitivity is constant within a region of interest (ROI). Pooling over the $N_{\text{ROI}}$ voxels (indexed by $p$ below) within the ROI $\mathcal{N}(r)$ centred on a target location, $r$, the elements of the outer product are formed from the inner product of the spatial signals measured on coil-wise pairs, indexed by $j = 1..N_{\text{coil}}$ and $k = 1..N_{\text{coil}}$:

$$\hat{E}_{r,j,k} = \sum_{p \in \mathcal{N}(r)} C_{j,p} C_{k,p}^* \quad (2)$$

In this case, $\widehat{\boldsymbol{E}}_r$ is no longer rank 1 due to noise, though the first singular vector will continue to explain the majority of the outer product variance, and to be a scaled version of the relative coil sensitivities. In practice, the tensor $\widehat{\mathbf{E}} \in \mathbb{C}^{N_{\text{voxel}} \times N_{\text{coil}} \times N_{\text{coil}}}$ can be obtained by convolving the tensor $\mathbf{E}$ with a smoothing kernel of size $N_{\text{ROI}}$ in image space.

### C. MORSE

The following sections build on the method of Walsh *et al.*, outlining a robust and computationally efficient means of estimating relative coil sensitivities and using these to unfold aliased images in a regularised SENSE framework. Computational efficiency was crucial to deployment for real-time use at the scanner console motivating many of the implementation choices.

#### 1) Reduced voxel-wise computation

MORSE operates in a reduced dimensionality ("virtual coil") space. An initial SVD on the array coil data [19-22] extracts the principal modes of variation of the coils across space. Due to the linearity of the Fourier transform this step can be performed in k-space. Typical receiver arrays have coil elements with spatially localised regions of high sensitivity. However, the virtual coils require fewer coils, $N_{\text{ref}}$, to achieve non-zero magnitude support over the entire imaging FoV. We therefore compute a reduced form outer product $\boldsymbol{E}_r^{\text{v}} \in \mathbb{C}^{N_{\text{coil}} \times N_{\text{ref}}}$ where $N_{\text{ref}} \leq N_{\text{coil}}$, and the superscript v denotes that we are working with virtual coils. This computationally efficient approach reduces $N_{\text{coil}}^2$ fitting problems per voxel to $N_{\text{coil}}$ x $N_{\text{ref}}$.

#### 2) Flexible spatial weighting

In MORSE, the uniform weights of Walsh *et al.* are replaced by arbitrary weights, $\boldsymbol{g} \in \mathbb{R}^{N_{\text{ROI}}}$:

$$\widehat{\mathbf{E}}^{\text{v}} = \mathbf{E}^{\text{v}} * \boldsymbol{g} \quad (3)$$

For example, a 3D Gaussian kernel, centred on location $r$, up-weights the sensitivity estimates near this voxel and down-weights those further away. This is particularly beneficial at the periphery of the imaged anatomy because it facilitates a degree of extrapolation of the sensitivity estimates that affords robustness to motion between the fully sampled reference data and the under-sampled images, if acquired separately. For computational efficiency, this convolution in image space is implemented as a multiplication in k-space.

#### 3) Higher order sensitivity estimation

The equivalent SVD to (1) is performed voxel-wise on the weighted outer product, $\widehat{\boldsymbol{E}}_r^{\text{v}} \in \mathbb{C}^{N_{\text{coil}} \times N_{\text{ref}}}$:

$$\widehat{\boldsymbol{E}}_r^{\text{v}} = \boldsymbol{U}_r \boldsymbol{S}_r \boldsymbol{V}_r^H \quad (4)$$

The left singular vectors, i.e. columns of $\boldsymbol{U}_r$, capture the contribution of the coil sensitivities relative to the dominant virtual coil mode. $\widehat{\boldsymbol{E}}_r^{\text{v}}$ will have rank $> 1$, even in the absence of noise, in scenarios such as when:

1) sensitivities vary over a spatial scale smaller than that of the spatial filter, $\boldsymbol{g}$, which is more likely at UHF;
2) the FoV is insufficient leading to spatial aliasing in the reference data;
3) chemical shift artefact induces spatial mis-localisation of the MRI signal.

In such cases, $\boldsymbol{S}_r$ contains multiple non-zero singular values quantifying the degree to which the corresponding left singular vectors (columns of $\boldsymbol{U}_r$) capture the modes of relative sensitivity variation of the virtual coils. Although the higher-order sensitivities will be orthogonal at a given location, this may not be the case across aliased voxel locations. Analogous to the approach used in ESPIRiT [3], retaining multiple "higher-order" sensitivity estimates per voxel, i.e. $k = 1..N_{\text{order}}$ where $1 < N_{\text{order}} \leq N_{\text{ref}}$ should increase the robustness to intra-ROI sensitivity variation thereby improving subsequent unfolding performance. The final coil sensitivity estimates, $\widehat{\mathbf{B}}^{\text{v}} \in \mathbb{C}^{N_{\text{coil}} \times N_{\text{order}} \times N_{\text{voxels}}}$, are formed voxel-wise by retaining the first $N_{\text{order}}$ columns of $\boldsymbol{U}_r$ from (4):

$$\hat{B}^{\text{v}}_{j,k,r} = U_{j,k,r} \; \forall \; k \leq N_{\text{order}} \quad (5)$$

As a final step, for consistency with the aliased images to be unfolded, the sensitivity estimates are transformed from the virtual to the original coil space to give $\hat{\boldsymbol{B}}$. Doing so is particularly computationally efficient for time series data (e.g. in fMRI context) since the transformation only has to be performed once (for the sensitivity estimates) rather than for each under-sampled volume.

### 4) Unfolding in a regularised SENSE framework

Under-sampling leads to spatial aliasing of the magnetisation distribution of interest with a pattern determined by the acceleration factor and k-space sampling scheme of the acquisition. At a given location, the coil-wise, aliased signals, $\boldsymbol{a} \in \mathbb{C}^{N_{\text{coil}} \times 1}$, are the superposition of the underlying magnetisation, $\boldsymbol{\rho} \in \mathbb{C}^{N_{\text{alias}} \times 1}$ weighted by the coil sensitivities at the origin locations:

$$\boldsymbol{a} = \boldsymbol{X}\boldsymbol{\rho} + \boldsymbol{\varepsilon} \quad (6)$$

Here, $\boldsymbol{X} \in \mathbb{C}^{N_{\text{coil}} \times N_{\text{alias}}}$ contains the coil-wise sensitivity estimates at the set of $N_{\text{alias}}$ voxel locations, $\mathcal{A}(r)$, and $\boldsymbol{\varepsilon} \in \mathbb{C}^{N_{\text{coil}} \times 1}$ is the measurement noise. Higher order sensitivity estimates at each aliased voxel location are incorporated as additional columns in $\boldsymbol{X}$, which is expanded to $\boldsymbol{X} \in \mathbb{C}^{N_{\text{coil}} \times (N_{\text{alias}} N_{\text{order}})}$ such that for coil $j$, the corresponding row is:

$$\boldsymbol{x}_j = \left[ \{\hat{B}_{j,1,r}\}_{r \in \mathcal{A}} \; \ldots \; \{\hat{B}_{j,N_{\text{order}},r}\}_{r \in \mathcal{A}} \right] \quad (7)$$

When inverting this expanded forward model to recover the underlying magnetisation, Tikhonov regularisation can be used to give a more noise-robust solution [23]. The singular values (diagonal elements of $\boldsymbol{S}_r$) from (4) will only be appreciable where there is non-zero magnetisation. Combined with an additional global scalar, $\lambda \in \mathbb{R}_+$, the singular values can be used to control the trade-off between confidence in the sensitivity estimates and the goodness of fit of the reconstructed image to the measured coil-wise images. The regularisation terms, $\left\{ \frac{\lambda}{S_{r,k,k}} \right\}_{r \in \mathcal{A}}$, specific to the sensitivity order (indexed by $k$) and alias locations, are assembled in a diagonal matrix, $\boldsymbol{\Lambda} \in \mathbb{C}^{N_{\text{alias}} N_{\text{order}} \times N_{\text{alias}} N_{\text{order}}}$. This augments the SENSE formalism to estimate the regularised solution for the magnetisation distribution:

$$\hat{\boldsymbol{\rho}} = (\boldsymbol{X}^H \boldsymbol{X} + \boldsymbol{\Lambda})^{-1} \boldsymbol{X}^H \boldsymbol{a} \quad (8)$$

$\hat{\boldsymbol{\rho}}$ would have length $N_{\text{alias}} N_{\text{order}}$, however, for computational efficiency, the pseudo-inverse in (8) is truncated to retain only the first $N_{\text{alias}}$ rows to estimate the first order unaliased signal $\hat{\boldsymbol{\rho}} \in \mathbb{C}^{N_{\text{alias}}}$.

## III. METHODS

The MORSE image reconstruction scheme has been deployed for use in multiple 3T and 7T functional and structural imaging studies conducted at The Functional Imaging Laboratory (The FIL) within UCL's Department of Imaging Neuroscience. As such it has been tested in many hundreds of participants, including healthy participants and diverse patient cohorts, with a variety of acquisition protocols. In the spirit of reproducible research, the open-source code, as well as scripts necessary to compile a Gadgetron image and create a Docker container encapsulating the MORSE reconstruction, are provided via GitHub. In what follows we provide an overview of the implementation and evaluate performance via exemplar *in vivo* data.

### A. Gadgetron and MATLAB implementation

The MORSE implementation is deployed for real-time image reconstruction as a MATLAB (Mathworks, Natick, MA) gadget within the Gadgetron image reconstruction framework [24]. MATLAB was used as the development environment to facilitate rapid prototyping. Native Gadgetron gadgets pre-whiten the data and extract the reference and under-sampled data into separate data structures. For structural imaging, readout oversampling is removed within Gadgetron prior to sending the data to MATLAB.

Within MATLAB, the fully sampled reference data are apodised, using a Tukey filter, to reduce Gibbs ringing. Next, the data are transformed into virtual coil space by performing an SVD over the entire spatial coverage having reformatted the data into a voxels-by-coils matrix. In virtual coil space, the data are zero-filled, if necessary, to the target resolution of the under-sampled images to be unfolded, and transformed to the image domain via inverse Fast Fourier Transform. The reduced outer product of the virtual coils is formed for each voxel (i.e. $\boldsymbol{E}^{\text{v}}_r \in \mathbb{C}^{N_{\text{coil}} \times N_{\text{ref}}}$). Multiplication in k-space is used to efficiently combine these outer product matrices over voxels (3) with a Gaussian filter parameterised by its full width at half maximum, $w$, defined in voxels. Subsequently, an (economy/reduced) SVD (4) is calculated voxel-wise. For computational efficiency, this is done for all voxels in a single operation using the 'pagesvd' command in MATLAB. The resulting sensitivity estimates (5) are returned to the original coil space where the pseudo-inverse (8) is computed for use in a regularised SENSE reconstruction of the under-sampled data.

### B. Exemplar in vivo evaluation

MORSE was assessed using a variety of data acquired with approval from local research ethics committees, and following informed consent. The 3T data presented here were acquired on a Siemens Prisma$^{\text{fit}}$ (VE11c software version) using a body coil for transmission and a 64-channel head and neck array for reception. The 7T data were acquired on a Siemens Terra (VE12u-SP01 software version) using an 8-channel transmit, 32-channel receive Nova Medical head coil operating in a quadrature-like ("TrueForm") mode. Key protocol parameters of all exemplar acquisitions are provided in Table I.

### C. Effect of varying reconstruction parameters

A parameter sweep approach was used to assess the impact of the free parameters within the MORSE image reconstruction pipeline. fMRI data were acquired from three healthy participants volunteering in a study of autobiographical memory encoding at 7T using a T2*-weighted 3D-EPI readout. Reference data were acquired as a fully sampled pre-scan with matched readout [25]. For each parameter set, the image quality of the EPI time series (approximately 200 volumes) was assessed qualitatively via visual inspection and quantitatively via the temporal signal-to-noise ratio (tSNR). The tSNR was summarised by the minimum, maximum and mean values within an ROI in frontal cortex, which demonstrated vulnerability to $B_o$ inhomogeneity, as well as suffering from residual aliasing artefacts due to the constrained anterior-posterior FoV. The image reconstruction parameters were varied as follows: $N_{order}$ = [1,2,3,4], $N_{ref}$ = [2,4,6,8], $w$ = [0.1,1,3,5,7,9,11,13,15] voxels and $\lambda$ = [0.001, 0.01, 0.1, 0.2, 0.3, 0.4, 0.5, 0.6, 0.7, 0.8, 0.9, 1, 10]. Note that $\lambda$ depends on the amplitude of the k-space data meaning that the appropriate value range will depend on a variety of factors and be application-specific. The parameters were incremented with a variable step size to demonstrate the effect of reconstructing using both extreme (sparsely sampled) and near-optimal (densely sampled) values. While a given parameter was swept, the others were kept fixed at $N_{order}$ = 2, $N_{ref}$ = 8, $w$ = 5 and $\lambda$ = 0.1. The tSNR was calculated for each of the resulting fMRI time series following motion correction using SPM12 and high pass filtering with a cut-off frequency of 1/128 Hz.

### D. Comparison with alternative image reconstruction schemes

MORSE was compared with the vendor-provided GRAPPA reconstruction [2, 15] using default parameters, with ESPIRiT [3] from the BART toolbox v0.4.04, and with LORAKS v2.1. ESPIRiT was implemented by calling the "ecalib" function for coil sensitivity estimation and "pics" for unfolding, both with default settings. We hypothesized that including $m$ coil sensitivities in the "Soft SENSE" reconstruction formalism [3] would have a similar effect to using $N_{order} \approx m$ in MORSE and therefore tested $m$ = [1,2] for functional and $m$ = [1,2,4] for structural imaging. LORAKS was called using the "AC_LORAKS.m" function with matrix type C, $\lambda$ = 0.5, rank 150 and Algorithm 4 [18].

Task-free 3D-EPI time series data (50 volumes) were acquired on three further participants at 7T with 4 (in-plane) x 2 (through-slab) acceleration and reconstructed with MORSE using the fixed values of the parameter sweep evaluation ($N_{order}$ = 2, $N_{ref}$ = 8, $w$ = 5 voxels and $\lambda$ = 0.1). The mean of the time series was segmented using SPM12 to identify grey and white matter voxels. A liberal brain mask was defined by those voxels for which the joint probability of being in grey or white matter exceeded 50% and was computed for each reconstruction separately. To avoid bias towards any particular reconstruction the tSNR was summarised by extracting the mean value from the logical conjunction of these masks.

Structural images were acquired at 3T and 7T using an in-house but openly-available 3D multi-echo spoiled gradient echo acquisition configured with either proton density (PD), T1 or magnetisation transfer (MT) weighting. An under-sampling factor of two was used in each phase-encoded direction, with integrated reference data. To further ease replication at other sites, additional data was acquired at 7T with a vendor-provided (i.e. "product") sequence. For all structural data, the MORSE reconstruction used $N_{ref}$ = 6, $w$ = 6 voxels, $N_{order}$ = [1,4,6] and $\lambda$ = $3\times10^{-4}$ as the global regularisation factor. Full details are provided in Table I.

The computational efficiency of each reconstruction was assessed using these structural data. For consistency, this was done offline with the identical k-space data. The MORSE and ESPIRiT reconstructions were performed on a Dell 7920 with Intel Xeon Gold 6230R 2.1GHz Processor and 768 GB of RAM (equivalent to what is used at The FIL for online MORSE reconstruction), while the offline GRAPPA reconstruction used the native scanner hardware via the "retrospective reconstruction" facility.

TABLE I
KEY SEQUENCE PARAMETERS USED FOR THE EXEMPLAR RECONSTRUCTIONS.
TE: ECHO TIME. TR: REPETITION TIME. TA: ACQUISITION TIME. HF: HEAD-TO-FOOT. AP: ANTERIOR-POSTERIOR. RL: RIGHT-TO-LEFT.

| Field [T] | Resolution [mm³] | Acceleration | Separate reference data | Matrix [HF x AP x RL] | TR [ms] | TE [ms] | TA [s/vol] |
|---|---|---|---|---|---|---|---|
| *Functional Imaging: 3D-EPI* | | | | | | | |
| 7 | 0.8 x 0.8 x 0.8 | 4 (PE) x 2 (3D) | 88 x 240 x 240 | 88 x 240 x 240 | 43.0 | 18.5 | 3.8 |
| *Parameter sweep analysis:* 3 participants, autobiographical memory task, 2 runs per participant (c. 200 volumes/run). *Image reconstruction comparison:* 50 task-free volumes from 3 further participants. | | | | | | | |
| 3 | 3.0 x 3.0 x 3.0 | 1 (PE) x 2 (3D) | 48 x 64 x 64 | 48 x 64 x 64 | 44.0 | 22.2 | 2.1 |
| Single participant with CAIPI sampling trajectory. | | | | | | | |

| Field [T] | Resolution [mm³] | Acceleration | Integrated reference data | Matrix [HF X AP x RL] | TR [ms] | TE,ΔTE [ms] | TA [min:s] |
|---|---|---|---|---|---|---|---|
| *Structural Imaging: Gradient echo* | | | | | | | |
| 3 | 0.8 x 0.8 x 0.8 | 2 (PE) x 2 (3D) | 40 x 40 lines | 320 x 280 x 224 | 44.0 | 2.3, 2.3 | 7:08 |
| PD- or T1-weighting was achieved with a flip angle of 6° or 21° respectively, 8 echoes. MT-weighting was induced by adding an off-resonance pre-pulse prior to the PD-weighted acquisition (at both 3T and 7T). | | | | | | | |
| 7 | 1.5 x 1.1 x 1.1 | 2 (PE) x 2 (3D) | 48 x 48 lines | 72 x 200 x 200 | 21.0 | 9.5, 4.9 | 1:37 |
| Vendor SWI sequence "t2_swi_tr_p3_fast" found under SIEMENS\head\library\general on VE12u-SP01. The resolution was reduced, a second echo was added, interpolation was removed and the total acceleration was increased to a factor of 4. | | | | | | | |
| 7 | 0.6 x 0.6 x 0.6 | 2 (PE) x 2 (3D) | 48 x 48 lines | 426 x 364 x 288 | 19.5 | 2.2, 2.4 | 9:10 |
| PD- or T1-weighting was achieved with a flip angle of 6° or 24° respectively, 6 echoes. The PD-weighted protocol was additionally run with the vendor-provided sequence with a TR of 20.0 ms, leading to a TA of 9:15. | | | | | | | |

## IV. RESULTS

### A. Effect of varying reconstruction parameters

Fig. 1 illustrates the impact of varying the free parameters of the MORSE reconstruction on the tSNR of 7T 3D-EPI data from one exemplar individual. Following visual assessment, the quantitative analysis focussed on a small region of the inferior frontal cortex where residual aliasing artefacts were observed due to the constrained anterior-posterior field of view (Fig. 1a). As $\lambda$ increased there was initially a rapid increase in tSNR as the regularisation suppressed noise (Fig. 1b). For intermediate values the tSNR plateaued. At the highest values of $\lambda$ the tSNR was more spatially variable because noise suppression was accompanied by signal biases due to residual aliasing.

As the sensitivity order, $N_{\text{order}}$ (Fig. 1c), increased from 1 to 2 the maximum and mean tSNR decreased substantially. However, this was driven by a reduction in signal because unfolding efficacy improved and the signals from multiple spatial locations were no longer superimposed. Increasing the sensitivity order further had little impact on the tSNR but increased the computational overhead.

The tSNR showed minimal dependence on the number of reference coils, $N_{\text{ref}}$ (Fig. 1d), suggesting that two references are sufficient to provide support over this region of interest. However, this will not necessarily be the case over a larger spatial extent than the ROI considered in this analysis.

When the width of the smoothing kernel, $w$ (Fig. 1e), was small (0.1 voxels), effectively no smoothing was applied. In this case, the tSNR map showed a high degree of correspondence with the result for $N_{\text{order}} = 1$, even though the sensitivity order was in fact 2. This can be attributed to the fact that, without smoothing, the outer product matrix ($\widehat{E}_r^v$) was still rank 1 allowing only a single sensitivity to be estimated per voxel. As the width of the smoothing kernel increased, effectively increasing the ROI over which the sensitivity estimates were made, the rank of the matrix increased allowing estimation of the higher-order sensitivity, and an improvement in unfolding, manifesting as a reduction in tSNR. With a kernel width of a few voxels, the tSNR homogenised spatially with the maximum decreasing and the minimum increasing. Larger kernel widths ($w > 10$ voxels) led to more spatially variable tSNR with the maximum and minimum increasing and decreasing respectively.

Overall, this parameter sweep revealed that the tSNR was comparatively insensitive to each parameter of the reconstruction over a reasonably broad range of values, but that the optimal values were spatially-specific. These findings were consistent across each of the three individuals, though the exact spatial distribution of higher or lower tSNR was participant-specific.

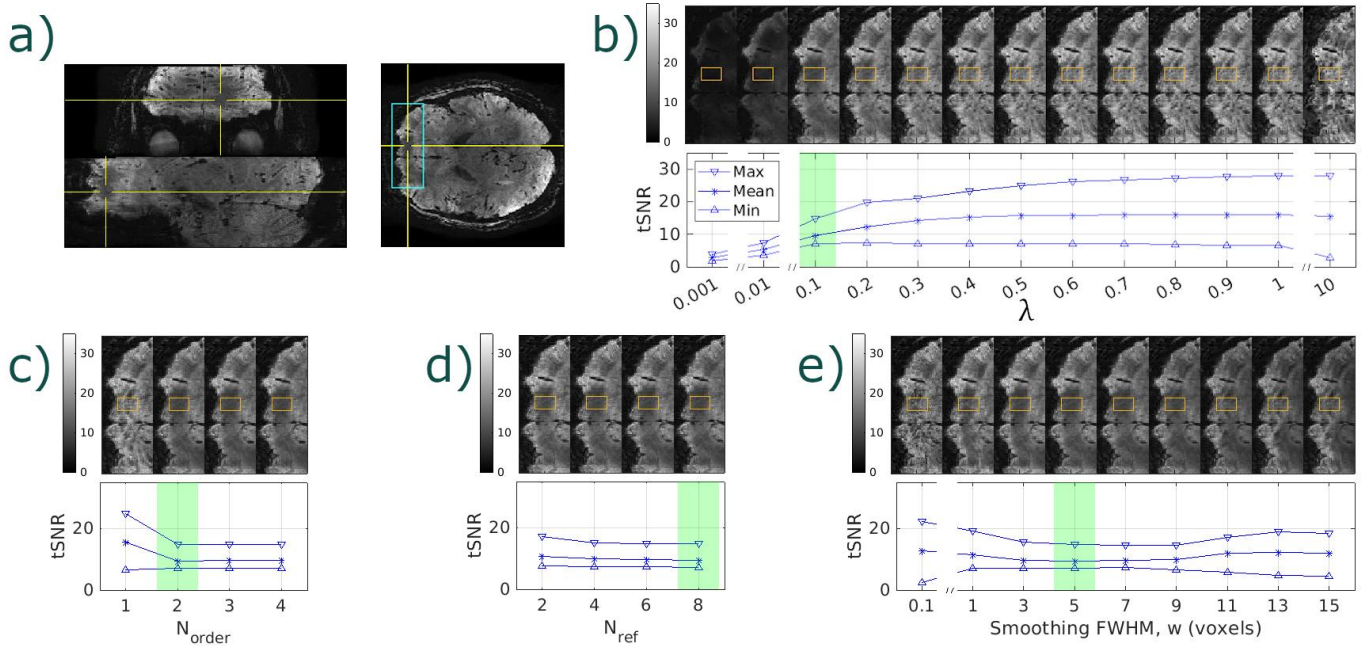

Fig. 1. tSNR as a function of the free parameters of the MORSE image reconstruction. tSNR map shown in a) with the yellow crosshairs highlighting a region of residual artefact, enclosed in a cyan box. This defines the anatomical extent displayed in sub-panels b-e. These sub-panels show the tSNR maps for this region, and the summary statistics (mean, min, max) extracted from the vicinity of the residual artefact (orange box) as the global regularisation parameter $\lambda$, b), the sensitivity order $N_{\text{order}}$, c), number of reference coils $N_{\text{ref}}$, d) or smoothing kernel width $w$, e) were varied. As a given parameter varied, the others were fixed at the values highlighted in green, which were also used to produce the tSNR maps in a).

### B. Comparison with alternative image reconstruction schemes

Fig. 2 illustrates the relative performance of the GRAPPA, ESPIRiT and MORSE reconstructions for a T2*-weighted EPI time series at 7T, via the tSNR. The GRAPPA reconstruction suffered from focal regions of rapidly varying signal intensity in frontal and temporal cortices in the proximity of increased $B_0$ field inhomogeneity (blue arrows). The ESPIRiT reconstruction with $m=1$ suffered from residual fold-over artefact that reduced the tSNR (red arrows) and sharp transitions between regions of low and high tSNR (orange arrows). No foldover artefact was visible in the mean EPI (not shown) or the tSNR map reconstructed using MORSE. The MORSE reconstruction also had the most homogenous tSNR distribution and the highest mean values for all three participants: tSNR$_{GRAPPA}$ = {12.11, 9.77, 11.28}, tSNR$_{ESPIRiT}$={9.73, 8.65, 10.32}, and tSNR$_{MORSE}$={13.55, 11.00, 12.87}. Increasing $m$ from 1 to 2 in ESPIRiT substantially degraded tSNR (data not shown; mean tSNR for the participant shown in Fig. 2 was reduced to just 4.64).

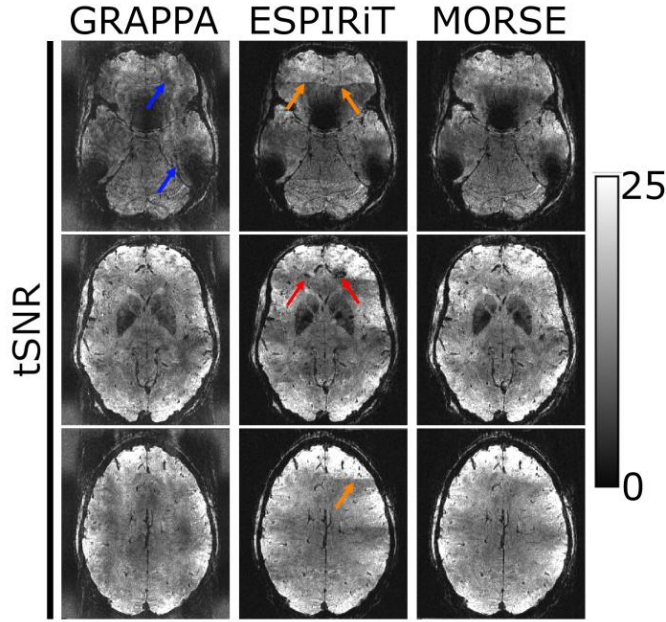

Fig. 2. tSNR maps for GRAPPA, ESPIRiT with m=1 and the proposed MORSE reconstruction with $N_{order}$=2. Image quality and tSNR (c.f. orange and red arrows) were consistently higher for the MORSE reconstruction.

Fig. 3 illustrates the relative performance of the GRAPPA, ESPIRiT and MORSE reconstructions for a T1-weighted gradient echo image at 3T. In this particular dataset, the final FoV (i.e. after unfolding) in the right-left and anterior-posterior directions was insufficient to encompass the full extent of the ears and the neck respectively. The GRAPPA, ESPIRiT with $m = 1$, and MORSE with $N_{order} = 1$ reconstructions failed to fully compensate for this leading to residual aliasing of the ears within the brainstem (Fig. 3, red box). Increasing the sensitivity order for MORSE ($N_{order} > 1$) or ESPIRiT ($m > 1$) improved the quality of the image reconstruction. However, for ESPIRiT this was at the cost of noise amplification despite residual artefact remaining even with $m = 4$. With $N_{order} = 6$, the artefact was minimised in the MORSE reconstruction with negligible noise amplification. These observations were independent of image contrast (we also analysed PD- and MT-weighted images from the same individual, results not shown).

The ESPIRiT, MORSE and GRAPPA reconstructions took approximatively 2.5 hours, 5.5 minutes and 3.5 minutes respectively for this protocol with 64 receive channels, a matrix size of 320x280x224 and 8 echoes.

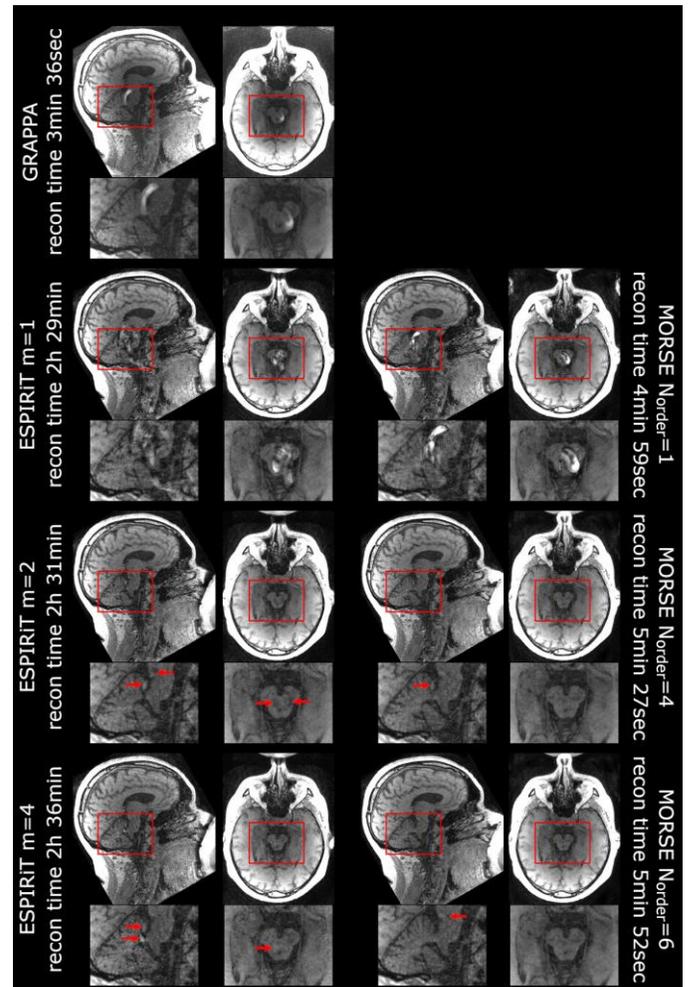

Fig. 3. T1-weighted data (first echo) reconstructed with GRAPPA, ESPIRiT and MORSE. The red boxes highlight a region with residual foldover artefacts, shown zoomed below the corresponding view with red arrows highlighting the more subtle artefacts. Higher-order sensitivities reduce the artefact for the MORSE and ESPIRiT reconstructions, but with less noise amplification and a significantly (> 25-fold) shorter reconstruction time for MORSE.

Fig. 4 illustrates the relative performance of the GRAPPA, ESPIRiT and MORSE reconstructions for a PD-weighted gradient echo image at 7T. The coil sensitivity varies rapidly at the periphery of the head (yellow arrows), necessitating higher-order sensitivities. Residual fold-over artefact was observed in the GRAPPA, ESPIRiT with $m = 1$ and MORSE with $N_{order} = 1$ reconstructions (red arrows). Increasing the sensitivity order for MORSE ($N_{order} > 1$) or ESPIRiT ($m > 1$) improved image quality. However, for ESPIRiT this was at the cost of noise amplification (Fig. 4, green circles) and some residual artefacts remaining even with $m = 4$. For $m > 4$, the noise amplification was excessive (data not shown). With $N_{order}$ of 4 or 6, the artefact was minimised in the MORSE reconstruction, with negligible noise amplification. These observations were independent of image contrast (we also analysed T1- and MT-weighted images from the same individual, results not shown).

The ESPIRiT, MORSE and GRAPPA reconstructions took approximatively 1 hour, 7.5 minutes and 1.5 minutes respectively for this protocol with 32 receive channels, a matrix size of 426x364x288 and 6 echoes.

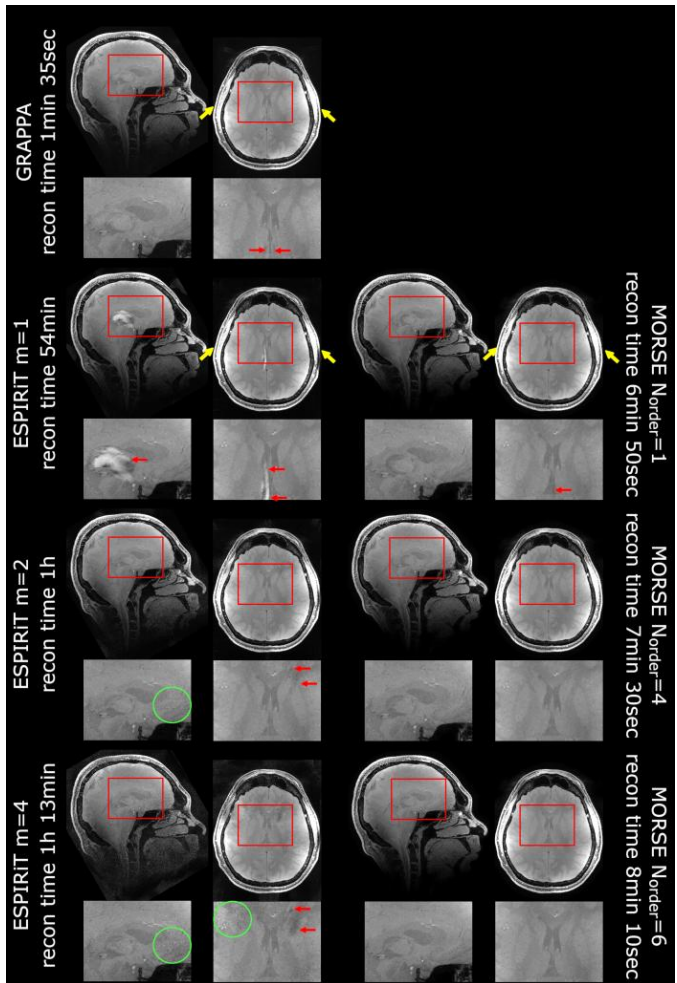

Fig. 4. 7T PD-weighted data (first echo) reconstructed with GRAPPA, ESPIRiT, and MORSE. The red boxes highlight a region with residual foldover artefacts, shown zoomed below the corresponding view with arrows highlighting the artefact. As higher-order sensitivities are incorporated in the MORSE and ESPIRiT reconstructions the rapidly varying coil sensitivity at the periphery (yellow arrows) is better captured and artefacts are reduced, particularly for the MORSE reconstruction. The ESPIRiT reconstruction exhibits noise amplification, highlighted by the green circles. The MORSE reconstruction had improved noise performance and a substantially (approximately 8-fold) shorter reconstruction time.

The above data had high isotropic resolution, with 0.8mm and 0.6mm voxel lengths at 3T and 7T respectively, and were acquired using an in-house sequence. However, similar performance was seen with a vendor-provided susceptibility-weighted imaging sequence at 7T. Fig. 5 illustrates the relative performance of the GRAPPA, ESPIRiT and MORSE reconstructions of data acquired with this sequence using lower, anisotropic resolution of 1.5 x 1.1 x 1.1 mm³ than previous examples. The GRAPPA reconstruction was fastest with a 7 second reconstruction time, but was corrupted by residual foldover artefact (red arrows). The ESPIRiT with $m = 2$ and MORSE with $N_{\text{order}} = 4$ reconstructions were of similar quality with no foldover artefact. However, the ESPIRiT reconstruction took over 15 minutes to reconstruct, infeasible for real-time deployment. The MORSE reconstruction took just 24 seconds. When deployed online it finished very shortly after image acquisition because the coil sensitivity estimation starts part way through the acquisition, as soon as all of the reference data have been acquired. Using a vendor-provided isotropic high-resolution protocol similar to the in-house sequence yields the same conclusion: MORSE outperforms GRAPPA (image quality) and ESPIRiT (reconstruction time) (results not shown).

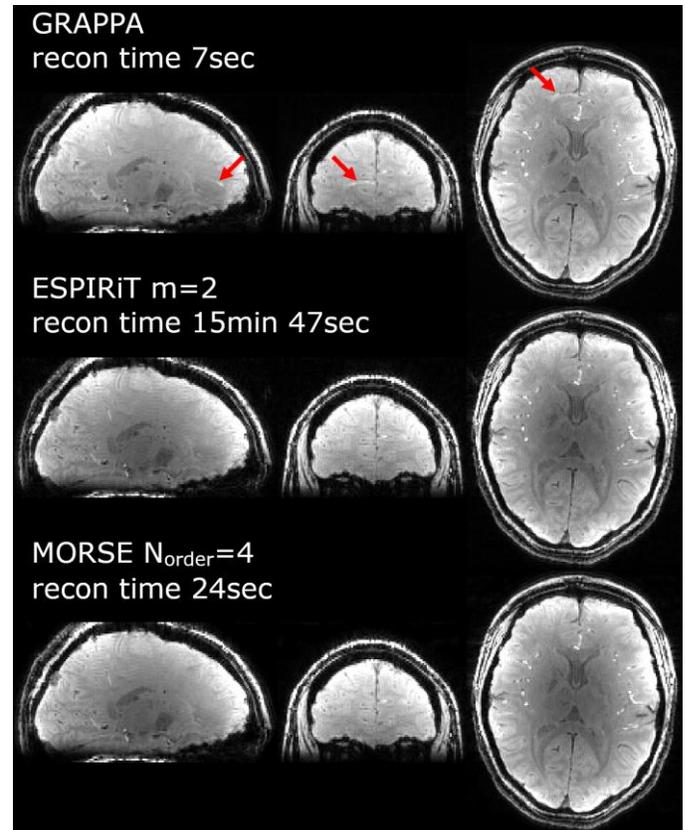

Fig. 5. GRAPPA, ESPIRiT, and MORSE reconstructions of a vendor-provided 3D GRE SWI sequence (first echo) with anisotropic 1.5 x 1.1 x 1.1 mm3 resolution at 7T. The GRAPPA reconstruction suffers from residual foldover artefact (red arrows) while the ESPIRiT and MORSE reconstructions produced images of comparable quality, free of foldover artefact. However, while ESPIRiT took 15min 47 seconds, MORSE took only 24 seconds. These data were acquired with a vendor-provided sequence - see Table I.

We additionally evaluated the LORAKS approach on high-resolution 7T PD- and T1-weighted data. At short echo times, LORAKS produced images comparable in quality to MORSE, free of fold-over artefacts and noise amplification, whereas the GRAPPA and ESPIRiT reconstructions were visibly affected by both. However, at later echo times, LORAKS consistently suffered from artefacts, as illustrated in Fig. 6 for the PD-weighted example. LORAKS took the longest to reconstruct (2 hours 10 minutes for the illustrated example).

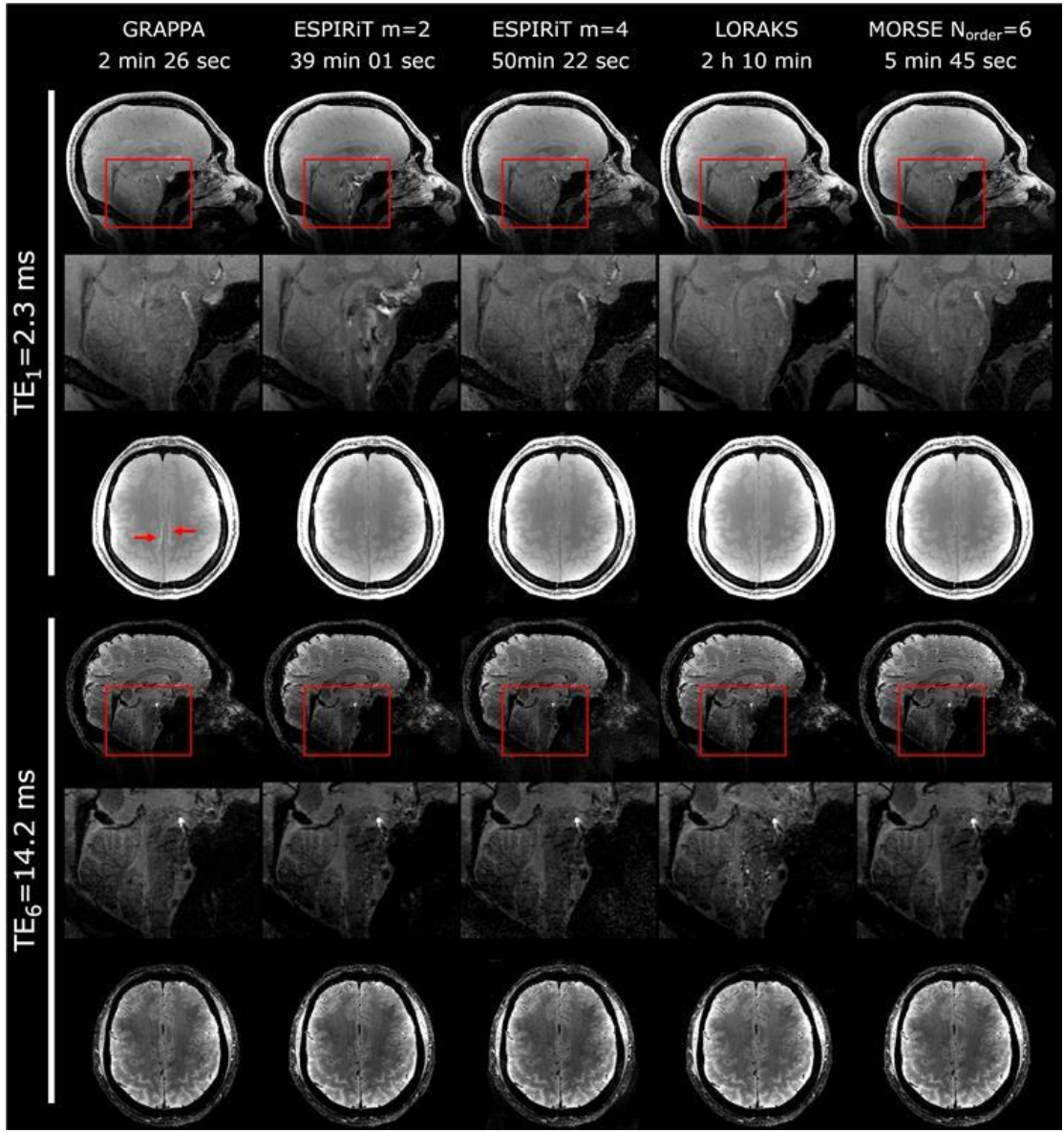

Fig. 6. 7T PD-weighted data (different participant to Fig. 4), reconstructed using GRAPPA, ESPIRiT, LORAKS, and MORSE. Red arrows indicate aliasing artefacts in the first echo of the GRAPPA reconstruction. Red boxes, and zoomed views of this region, highlight residual aliasing artefacts in the first echo of ESPIRiT and the last echo of LORAKS. The MORSE reconstruction yielded consistently high-quality images across all echoes, and had a reconstruction time commensurate with real-time deployment (5 minutes 45 seconds).

## V. DISCUSSION

MORSE is a flexible and computationally efficient method of estimating coil sensitivities together with a data-driven regularisation term for use in parallel imaging applications. MORSE robustly reconstructed under-sampled data producing images free of aliasing artefacts across multiple contrasts, readouts, sampling trajectories (CAIPIRINHA sampling in 3T 3D-EPI also tested, results now shown), resolutions and field strengths. MORSE has been successfully deployed in multiple neuroimaging studies at both 3T and 7T, including functional studies of autobiographical memory processing, visual and auditory perception, and quantitative MRI studies of neurodegenerative diseases such as Huntington's, Alzheimer's and Parkinson's.

MORSE delivered improved image quality relative to GRAPPA, ESPIRiT, and LORAKS in terms of both alias and noise suppression, with intermediate reconstruction times that were feasible for real-time deployment (Figs.2-6). The vendor-provided GRAPPA implementation offered the most computationally efficient reconstruction despite running on the least powerful hardware. However, it suffered from reduced tSNR (Fig. 2) and residual aliasing artefacts (Figs.3-6). The ESPIRiT implementation via the BART toolbox had one of the longest reconstruction times, which were not commensurate with real-time deployment. In 3D-EPI (Fig. 2), it consistently produced the lowest tSNR with abrupt changes in space that would translate into strong spatial variation in functional sensitivity. Residual aliasing artefacts were common with ESPIRiT, even when multiple eigenvectors (i.e. $m > 1$) were used. Although multiple eigenvectors reduced aliasing, this was at the cost of noise amplification. However, the ESPIRiT reconstructions used default settings, which may have been suboptimal. Efforts to modify these to improve the reconstructions, via coil compression, regularisation and eigenvalue threshold, were unsuccessful (data not shown) suggesting that the default settings were at least close to optimal. As with ESPIRiT, the reconstruction time of LORAKS was prohibitively long for real-time deployment, though the method is increasingly explored in research applications because of its flexibility. We systematically evaluated a range of algorithm settings (twelve additional parameter combinations beyond the optimal combination shown in Fig. 6) manipulating the data whitening, regularisation, matrix types C and S, Algorithms 3 and 4, and separate versus joint (across echoes) reconstructions [16]. These yielded results that were either comparable to or worse than the example presented in Fig. 6. Reconstruction times ranged from 2 hours 7 minutes to 31 hours 14 minutes.

The free parameters of the MORSE reconstruction proved robust to the choice of resolution, image contrast and field strength. In the diversity of datasets presented here the parameter sets differed only based on readout, i.e. for EPI or non-EPI data. The parameter sweep analysis (Fig. 1) showed that although the optimum combination of reconstruction settings is regionally specific, it has comparatively modest impact on tSNR over a broad range of parameter values. A key difference between the functional and structural data is the nature by which the reference data used to estimate the sensitivity data were acquired. In the case of 3D-EPI for fMRI, the reference data were acquired as a separate, fully sampled volume using segmentation to match readout and resolution. In the structural imaging examples, the reference data were acquired as an integrated, fully sampled reference region with substantially lower (~5mm voxel length) resolution than that of the under-sampled data to be reconstructed. The nature of the readout also means that the TE is about 10 times shorter for the structural imaging cases (see Table I). In our implementation, if multiple echoes are acquired, the shortest TE is used for coil sensitivity estimation by default to maximise SNR and minimise susceptibility-induced dephasing and dropouts. Use of longer echo times to estimate coil sensitivities lead to poorer image quality of the shorter echoes due to residual aliasing, and/or noise amplification depending on the value of the global regularisation factor (results not shown). This vulnerability may be exacerbated in structural imaging because of the low resolution of the reference data.

The inclusion of higher-order sensitivities addresses situations where multiple relative coil sensitivities can be used to describe a given voxel, e.g. due to a rapidly varying sensitivity field, an insufficient FoV, or a chemical shift. In these cases, the smoothed eigen outer product in (4) will contain columns that are different linear combinations of distinct coil sensitivities and will no longer be rank 1. $N_{\text{order}}$ of 2 proved sufficient for 3D-EPI, regardless of target resolution or field strength, and can benefit cases where imperfections in the water-selective excitation lead to residual fat signal, displaced due to its chemical shift, or if wrap-around occurs due to the tight FoV necessitated by the rapid imaging requirements of fMRI. The structural imaging examples benefited from even higher $N_{\text{order}}$ of 4 to 6. This may be due, at least in part, to the lower resolution of the integrated reference data used in these examples, but was particularly beneficial when the target FoV was insufficient to fully encompass the participant's head (e.g. Fig. 3).

Incorporating the data-driven regularisation term in (8) improves the g-factor of the final image reconstructions by providing a masking-like reduction of background voxels, which would only contribute noise. When $N_{\text{order}} > 1$, multiple singular values can be extracted from $S_r$ to construct order-specific regularisation terms. Therefore, higher-order sensitivity contributions can be inherently suppressed in voxels for which there is less evidence of these.

Online deployment and rapid prototyping were facilitated by implementing the reconstruction as a MATLAB gadget within the Gadgetron image reconstruction framework [24]. The requirement for robust online deployment in routine scanning makes reconstruction speed a key factor in optimal parameter selection. Aside from acquisition choices, the computational speed is dictated by the number of reference coils, $N_{\text{ref}}$, and the sensitivity order per voxel, $N_{\text{order}}$. $N_{\text{ref}}$ must be sufficiently high to ensure that in the virtual coil space there is support, i.e. non-zero reference sensitivity, in all anatomy-containing voxels within the imaging FoV. For typical receiver coils at 3T and 7T as few as six reference coils suffices. This offers a significant computational saving over the method of Walsh et al. [6] because the size of the outer product, which is computed voxel-wise, is reduced by a factor of 5 or more. The second SVD, which is performed on these outer-product matrices, captures the relative variance of the virtual coils over the spatial patch encompassed by the weighting function – in our implementation the Gaussian smoothing kernel. As noted previously, increasing $N_{\text{order}}$ improves unfolding performance, but it also increases the computational overhead. Therefore, for a given application, the minimum number that can produce artefact-free images should be selected. When using $N_{\text{ref}} = 6$ and $N_{\text{order}} = 4$ to reconstruct structural imaging data with 6 echoes, whole brain coverage and 0.6 mm isotropic resolution acquired on a 32-channel receiver coil with integrated reference data, MORSE can lead to a lag of up to 1-minute following completion of the data acquisition. This could be shortened by using faster hardware, e.g. GPU, leveraging greater coil compression in virtual coil space, or acquiring separate rather than integrated reference data though this runs the risk of increasing vulnerability to motion [5]. Separate reference data are acquired for fMRI applications using 3D-EPI leading to a lag at the outset of the acquisition. However, once the pseudo-inverse of the regularised sensitivities has been computed, the reconstruction amounts to a simple matrix multiplication and the reconstruction rapidly catches up with the data acquisition.

Use of the vendor-agnostic ISMRMRD data format [26] and Docker containers enables stable application-specific deployment, continuous development and eases replication across sites and computational/imaging platforms. The flexible nature of the Gadgetron implementation also allows additional quality assurance measures to be incorporated to further improve robustness. For example, the reference data can be assessed at the outset of each fMRI run and the run restarted if motion or other artefacts, which would be catastrophic for the ongoing time series, are observed.

## VI. CONCLUSIONS

MORSE is a means of estimating one or more sensitivities per voxel, together with a data-driven regularisation term, that can be combined in a regularised SENSE framework to deliver robust, high-quality reconstructions of a wide variety of under-sampled data. In all of the evaluated contexts, MORSE outperformed LORAKS, ESPIRiT and GRAPPA. MORSE was shown to operate well with a range of structural and functional neuroimaging protocols at 3T and 7T, with up to 8-fold acceleration. Given the feasibility of real-time reconstruction via Gadgetron, this open-source approach has been used extensively in our lab, including with diverse patient cohorts.

## VII. DATA AVAILABILITY STATEMENT

The open-source code, as well as scripts necessary to compile a Gadgetron image and create a Docker container encapsulating the MORSE reconstruction, are available at https://github.com/fil-physics/gadgetron-matlab. Example k-space datasets at 3T and 7T are available here: https://zenodo.org/records/13737462.

## ACKNOWLEDGMENT

This research was funded in whole, or in part, by the Wellcome Trust [203147/Z/16/Z and 226793/Z/22/Z]. For the purpose of Open Access, the author has applied a CC BY public copyright licence to any Author Accepted Manuscript version arising from this submission. We thank Christian Lambert and Charlotte Dore for sharing structural imaging data from the qMAP-PD study, as well as Eleanor A. Maguire and Yan Wu for sharing 7T fMRI data.